\documentclass[%
reprint,
superscriptaddress,
showpacs,
showkeys,
amsmath,amssymb,
aps,
]{revtex4-1}

\usepackage{graphicx}
\usepackage{dcolumn}
\usepackage{bm}
\usepackage{hyperref}
\usepackage{color}

\usepackage{chngcntr}

\begin{document}
	

	\title{LaN Structural and Topological Transitions Driven by Temperature and Pressure}
	
	\author{Wei-Chih Chen}
	\affiliation{Department of Physics, University of Alabama at Birmingham, Birmingham, Alabama 35294, USA}
	\author{Chia-Min Lin}
	\affiliation{Department of Physics, University of Alabama at Birmingham, Birmingham, Alabama 35294, USA}
	\author{Joseph Maciejko}
	\affiliation{Department of Physics, University of Alberta, Edmonton, AB T6G 2E1, Canada}
	\affiliation{Theoretical Physics Institute, University of Alberta, Edmonton, AB T6G 2E1, Canada}
	\author{Cheng-Chien Chen}
	\email{Corresponding author. Email address: chencc@uab.edu}
	\affiliation{Department of Physics, University of Alabama at Birmingham, Birmingham, Alabama 35294, USA}
		
	\date{\today}
	
	\begin{abstract}
	We study lanthanum mononitride LaN by first-principles calculations. The commonly reported rock-salt structure of $Fm\bar{3}m$ symmetry for rare-earth monopnictides is found to be dynamically unstable for LaN at zero temperature. Using density functional theory and evolutionary crystal prediction, we discover a new, dynamically stable structure with $P1$ symmetry at 0 K. This $P1$-LaN exhibits spontaneous electric polarization. Our {\it ab initio} molecular dynamics simulations of finite-temperature phonon spectra further suggest that LaN will undergo ferroelectric and structural transitions from $P1$ to $Fm\bar{3}m$ symmetry, when temperature is increased. Moreover, $P1$-LaN will transform to a tetragonal structure with $P4/nmm$ symmetry at a critical pressure $P=18$ GPa at 0 K. Electronic structures computed with an advanced hybrid functional show that the high-temperature rock-salt LaN can change from a trivial insulator to a strong topological insulator at $P \sim 14$ GPa. Together, our results indicate that when $P=14 - 18$ GPa, LaN can show simultaneous temperature-induced structural, ferroelectric, and topological transitions. Lanthanum monopnictides thereby provide a rich playground for exploring novel phases and phase transitions driven by temperature and pressure.
		
	\end{abstract}


	\keywords{lanthanum pnictides, topological insulator, ferroelectric material, temperature/pressure-driven phase transitions, density functional theory, molecular dynamics}
	
	
	\maketitle

	\section{\label{sec:level1}Introduction}
	
	Semimetallic lanthanum monopnictides such as LaAs, LaSb, and LaBi have attracted much attention because of their unusual extreme magnetoresistance due to electron-hole compensation~\cite{LaSb_XMR_tafti2015,LaBi_LaSb_topo_ARPES_niu2016,LaSb_LaBi_XMR_PNAS,LaAs_mBJ_XMR_yang2017}. These materials typically assume stable rock-salt structures, and they are promising for magnetic sensor and spintronics applications. Different from the above semimetals, LaP is found theoretically to be semiconducting with a narrow band gap ($< 0.1$ eV), and it displays high thermoelectric performance tunable by strain and pressure~\cite{LaP_1,LaP_2}. In addition to the intriguing magnetic and electronic behaviors of lanthanum monopnictides, researchers have recently paid attention to their non-trivial topological properties.
	 
In the single-particle picture, topological bulk materials can be roughly categorized into topological semimetals (TSMs) and topological insulators (TIs). TSMs are characterized by topologically protected band crossings~\cite{TSM}; they include nodal semimetals such as Weyl and Dirac semimetals~\cite{yan2017,armitage2018}, as well as more exotic varieties like nodal-line semimetals~\cite{fang2016}, type-II Weyl semimetals~\cite{soluyanov2015}, and multifold fermions~\cite{fang2012,wieder2016,bradlyn2016}. For insulators with the time-reversal symmetry (TRS), topology arises from band inversion at TR-invariant momenta caused by strong spin-orbit coupling and can be classified by the $\mathbb{Z}_2$ indices~\cite{TI_hasan2010,qi2011}. The topologically non-trivial phases exhibit robust helical surface states with spin-momentum locking, which forbids back-scattering and thus renders new potential applications in low-dissipation devices.
	 
	 LaX (X = N, P, As, Sb, and Bi) lacks conducting $f$-electrons and preserves TRS, and their band structures also show an energy gap between valence and conduction bands at each $k$-point in the Brillouin zone (BZ). Hence, LaX still permits the $\mathbb{Z}_2$ classifications, even though some lanthanum monopnictides are semimetallic. Experimentally, the signature of topologically nontrivial (bulk) insulating phases can be determined by the conducting surface states near high symmetry points in the surface BZ, and they can be observed directly by angle-resolved photoemission spectroscopy (ARPES). In ambient conditions, LaBi is confirmed to be topologically nontrivial by several experimental groups~\cite{LaBi_topo_wu2016, LaBi_topo_ncomms13942, LaBi_topo_ARPES_lou2017, LaBi_LaSb_topo_ARPES_niu2016, LaBi_topo_PRB_feng2018, LaX_arpes_nummy2018}. In contrast, LaAs and LaSb are found to be topologically trivial in ARPES experiments~\cite{LaSb_ARPES_zeng2016, LaBi_LaSb_topo_ARPES_niu2016, LaX_arpes_nummy2018}. Recently, topological phase transitions (TPTs) are especially intriguing, because of the possibility to control different topological phases~\cite{TPT_2012, monserrat2017antiferroelectric, fan2017transition, ZrTe5_2019,sie2019ultrafast, vaswani2020light, PhysRevB.103.115207}. In LaAs and LaSb, pressure-induced TPTs have been proposed~\cite{LaAs_hydrostatic_pressure, LaSb_topo_transition_HSE_guo2017}. Since LaN and LaP are expected to show similar crystal and electronic structures as other lanthanum monopnictides, TPTs also may be achieved by uniaxial strain or hydrostatic pressure.

		
	\begin{figure*}
		\includegraphics[width=\textwidth]{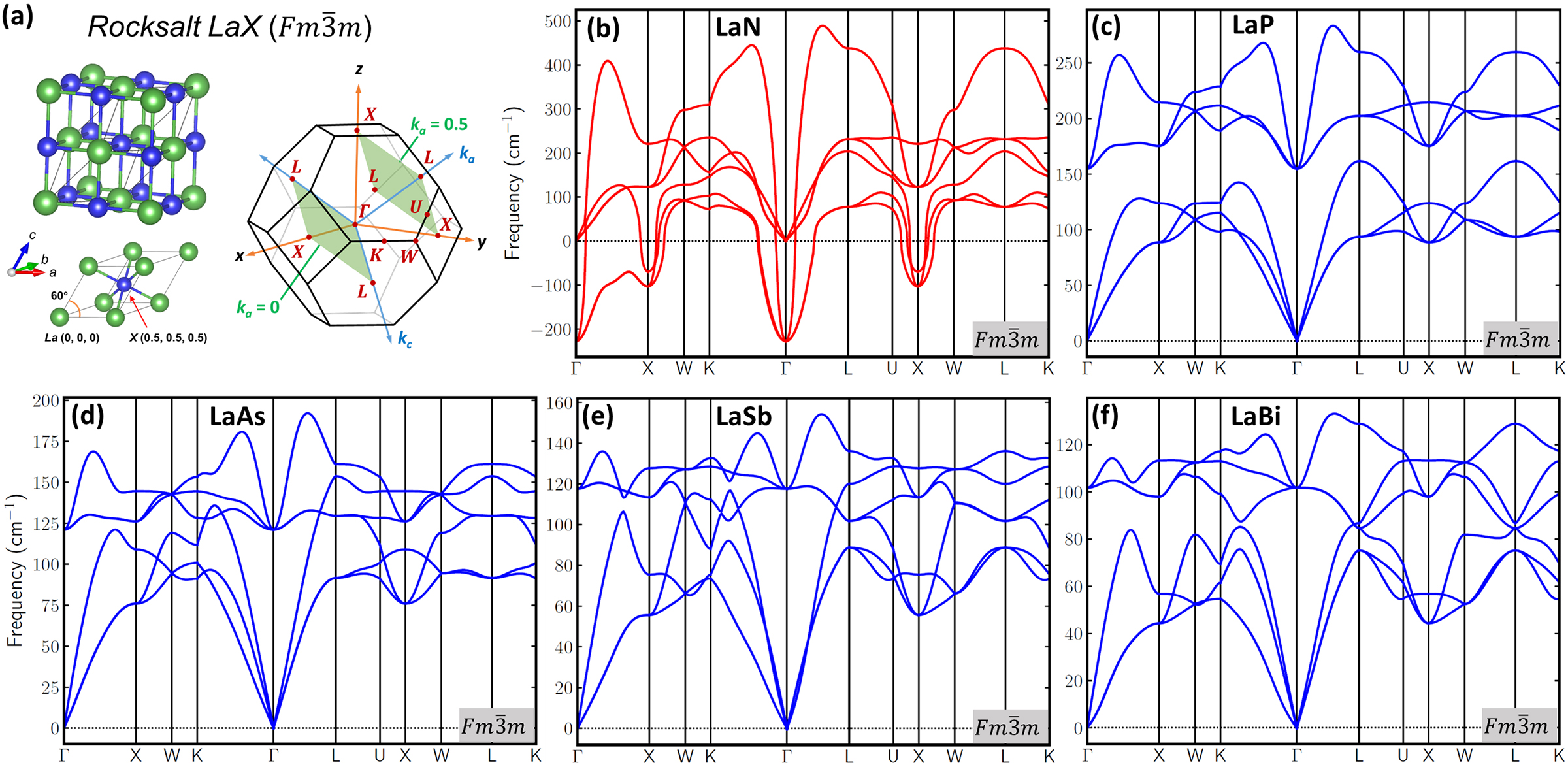}
		\caption{(a) Rock-salt structure of LaX (with X = N, P, As, Sb, or Bi), and the corresponding Brillouin zone with projected (001) and (111) surfaces. (b)-(f) Phonon dispersions (without non-analytical term correction) of LaN, LaP, LaAs, LaSb, and LaBi, respectively. The imaginary phonon modes in (b) show that the rock-salt LaN is dynamically unstable at 0 K.}
		\label{Fig.1}
	\end{figure*}
	
	In this paper, we use first-principles calculations to study the structural, electronic, and topological phase transitions of LaN. By re-examining the dynamic stabilities of all rock-salt LaX, we find that only LaN is dynamically unstable with imaginary phonon modes at zero temperature. Using evolutionary crystal structure prediction techniques, we find a new, dynamically stable structure of LaN with $P1$ symmetry. In our calculations, the $P1$-LaN also exhibits spontaneous electronic polarization, and it can undergo concomitant structural and ferroelectric transitions driven by temperature or pressure. 	The rest of the paper is organized as follows. In Sec. II, we discuss the computational methods based on density functional theory, evolutionary crystal structure prediction, and {\it ab initio} molecular dynamics. In Sec. III, we present the theoretical results of structural, electronic, and topological properties of LaN under different temperature and pressure. Sec. IV concludes the paper by summarizing the conditions for stabilizing different phases under study. The Appendix discusses additionally topological phase transitions in rock-salt LaP, LaAs, and LaSb under hydrostatic pressure.
	
	\section{\label{sec:level1}Methods}
	
	Our density functional theory (DFT)~\cite{DFT_1, DFT_2} calculations are conducted using the Vienna Ab initio Simulation Package (VASP)~\cite{VASP_1, VASP_2, VASP_GPU_1, VASP_GPU_2}. In particular, the Perdew-Burke-Ernzerhof generalized gradient approximation (PBE-GGA)~\cite{PBE} functional and the projector augmented wave (PAW) method~\cite{PAW_1, PAW_2} are adopted. The wavefunctions are expanded in plane-wave basis sets with a kinetic energy cutoff of 500 eV, and a $\Gamma$-centered 11 $\times$ 11 $\times$ 11  Monkhorst-Pack $k$ mesh~\cite{MP} is used for the Brillouin zone sampling. In structure relaxation calculations, the convergence criteria for the force on each atom and the total energy are set to $1\times 10^{-3}$ eV/\r{A} and $1 \times 10^{-8}$ eV, respectively.
	
		
	To search for the zero-temperature (0 K) crystal structure of LaN, the \text{USPEX} package~\cite{uspex_1, uspex_2} is used for crystal structure prediction (CSP). USPEX implements a highly efficient evolutionary algorithm for finding stable and meta-stable low-enthalpy structures corresponding to local minima in the potential energy surface. In our CSP calculations, 2-, 4-, and 8-atom unit cells are considered. The first generation of structures are randomly created. Subsequent generations are produced with 20\% from random structures and 80\% from heredity, soft mutation, and transmutation operators.
	
	The phonon dispersions at 0 K are calculated by the frozen-phonon approach with a $5\times 5 \times 5$ supercell (250 atoms) using VASP and PHONOPY~\cite{phonopy}. Here, VASP calculates the force on each atom in the supercell and provides the force information for PHONOPY, which in turn computes interatomic force constants and obtains phonon eigenvalues by diagonalizing the dynamical matrix. The atomic displacement is set as 0.02 \r{A}, and a 3 $\times$ 3 $\times$ 3 \textit{k}-mesh is employed in the self-consistency calculations. Since most of the rare-earth monopnictides are semimetallic, the non-analytical term correction is not included in our calculations.

	For finite-temperature phonon spectra, we adopt the temperature-dependent effective potential (TDEP)~\cite{TDEP} technique with \textit{ab initio} molecular dynamics (AIMD). The AIMD simulations use a time step of 2 fs and a total simulation length of 20 ps. We have checked the convergence of our results with respect to the total simulation time. The final 200 structures are then utilized to estimate the force constants via least-square fitting by using the ALAMODE package~\cite{Tadano_2014}.
	
	For electronic structure calculations, band orderings are essential for capturing the correct topological nature. However, standard DFT calculations based on local functionals like LDA or GGA typically underestimate the band gap, which may lead to an incorrect topological classification. To more adequately compute the band gap, we further apply an advanced non-local hybrid functional HSE06~\cite{HSE06} with spin-orbit coupling (SOC). Since hybrid DFT is computationally more demanding, we reduce the $k$ points to 9 $\times$ 9 $\times$ 9 and the cutoff energy to 400 eV (and we have checked that the band structure is quantitatively consistent with that obtained by using a 11 $\times$ 11 $\times$ 11 $k$ mesh and 500 eV energy cutoff). With the eigenvalues and eigenfunctions from hybrid-DFT, we construct maximally localized Wannier functions (MLWFs) and obtain corresponding \textit{ab-initio} tight-binding models using the Wannier90 package~\cite{wannier90}. We then use the WannierTools package~\cite{wanniertools} to identify topological $\mathbb{Z}_2$ indices by tracking the Wannier charge centers~\cite{z2pack}.

	\section{\label{sec:level1}Results and Discussion}

	Several rare-earth monopnictides are known to be stabilized in the cubic rock-salt structure (space group No. 255, $Fm\bar{3}m$ symmetry), as shown in Fig. 1(a). Indeed, a previous X-ray diffraction (XRD) experiment has reported the observation of rock-salt LaN~\cite{LaN_pressure_schneider2012}. However, a recent experiment using magnetron sputtering found only wurtzite and zinc blende structures for LaN~\cite{LaN_synthesis}. As a result, we begin by using DFT to examine the structural stability. We first fully relax the rock-salt structures for all lanthanum monopnictides using both LDA and GGA functionals without SOC. The resulting lattice parameters obtained by GGA (LDA) for LaX (with X ranging from N, P, As, Sb, to Bi) are 5.3076 (5.2140), 6.0496 (5.9377), 6.1925 (6.0656), 6.5466 (6.4019), and 6.6596 (6.5065) \r{A}, respectively. Compared with the experimental lattice parameters of 6.1670(7) \r{A} for LaAs~\cite{LaAs_mBJ_XMR_yang2017}, 6.5008 \r{A} for LaSb~\cite{LaSb_LaBi_XMR_PNAS}, and 6.5799 \r{A} for LaBi~\cite{LaSb_LaBi_XMR_PNAS}, GGA slightly overestimates the lattice parameters (with respective errors of 0.4\%, 0.7\%, and 1.2\%), while LDA underestimates them (with respective errors of -1.6\%, -1.5\%, and -1.1\%). We note that the SOC effect on lattice parameters is less than 0.1\%. Since overall the GGA results agree better with experiments, below we consider the fully relaxed structures obtained by GGA as the unstrained structures. 
	
	To examine the structural stabilities, we first compute the elastic constants using VASP~\cite{ELASTIC_VASP} to ensure that the Born-Huang criteria for cubic crystals are met~\cite{BORN-HUANG}: $C_{11}-C_{12} > 0$,  $C_{11}+2C_{12} > 0$, and $C_{44} > 0$. Our calculations indicate that all LaX rock-salt structures are mechanically stable. We next study the dynamical stability by calculating the phonon dispersion at 0 K. The phonon spectrum of LaN in Fig. 1(b) contains soft modes with imaginary frequencies around the $\Gamma$ and $X$ points, which indicates the dynamical instability of LaN's rock-salt structure. Our LaN phonon dispersion calculated by a finite-displacement method is consistent with the study of G\"ok\"o\u{g}lu \textit{et al.}~\cite{LaN_phonon} based on density functional perturbation theory. For other lanthanum monopnictides, the phonon spectra in Figs. 1(c)-(f) do not exhibit any soft mode, showing that the corresponding rock-salt structures are dynamically stable. This is also consistent with recent experiments~\cite{LaSb_ARPES_zeng2016, LaBi_LaSb_topo_ARPES_niu2016, LaX_arpes_nummy2018}. 
We further note that in our electronic structure calculation using the HSE06 hybrid functional~\cite{HSE06}, unstrained rock-salt LaP is a semiconductor with an indirect band gap $\sim 82$ meV. Therefore, a longitudinal optical-transverse optical (LO-TO) phonon splitting near $\Gamma$ may be observed experimentally.
	
	The imaginary phonon modes in rock-salt LaN imply a different stable crystal structure at 0 K. Using the evolutionary algorithm implemented in \text{USPEX}~\cite{uspex_1, uspex_2}, we conduct crystal structure searches at 0 K and find a lower-symmetry structure with space group No. 1 ($P1$ symmetry) shown in Fig. 2(a) as the ground state. The lattice parameters of the primitive cell are a = 3.7245 \r{A}, b = 3.8192 \r{A}, and c = 3.7557 \r{A}. The angles between them are $\alpha$ = 58.8990$^{\circ}$, $\beta$ = 61.3750$^{\circ}$, and $\gamma$ =   59.7033$^{\circ}$. The Wyckoff positions for La and N are (0, 0, 0) and (0.42, 0.52, 0.58), respectively. This structure can be regarded as a distorted rock-salt structure, where the N position is slightly shifted away from the more symmetric (0.5, 0.5, 0.5) position. Figure 2(b) shows the corresponding phonon dispersion, and the absence of imaginary mode indicates that $P1$-LaN is dynamically stable at 0 K.

	\begin{figure}
	\centering
	\includegraphics[width=\linewidth]{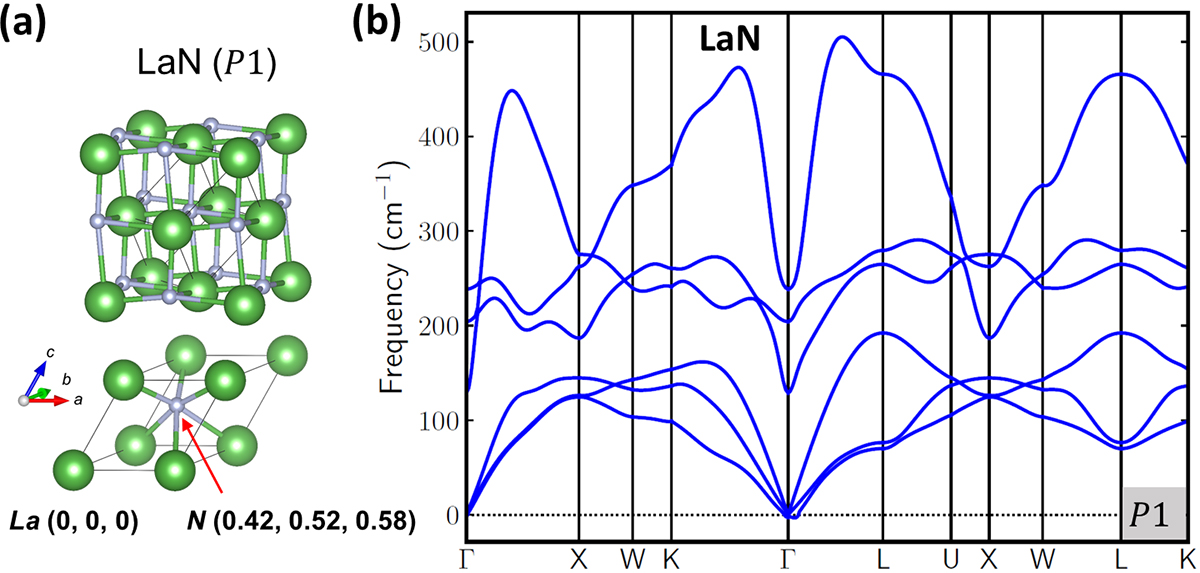}
	\caption {(a) Distorted rock-salt structure obtained by evolutionary algorithm for LaN. This structure lacks an inversion center and belongs to space group No. 1 (\textit{P1} symmetry). The lattice parameters are a = 3.7245 \r{A}, b = 3.8192 \r{A}, c = 3.7557 \r{A}, $\alpha$ = 58.8990$^{\circ}$, $\beta$ = 61.3750$^{\circ}$, and $\gamma$ = 59.7033$^{\circ}$. The Wyckoff positions are La (0, 0, 0) and Na (0.42, 0.52, 0.58). (b) Phonon dispersion of the $P1$-LaN structure. The labelings of high-symmetry points are based on the fcc lattice. The absence of imaginary phonon mode indicates that $P1$ LaN is dynamically stable at 0 K.}
	\label{Fig.2}
	\end{figure}

	To ensure that $P1$-LaN is indeed the ground state structure, we also compute the total energy versus volume for LaN in five different crystal structures: the low-symmetry structure $P1$ (space group No. 1), rock salt $Fm\bar{3}m$ (No. 225), PbO-like tetragonal structure $P4/nmm$ (No. 129), wurtzite $P6_3mc$ (No. 186), and zinc blende $Fm\bar{4}3m$ (No. 216). As shown in Fig. 3(a), the $P1$ structure has the lowest total energy, which is 29 meV (per formula unit of LaN) lower than that of the rock-salt structure. The lowest possible total energy of the wurtzite structure is also 16 meV (per formula unit) lower than that of the rock-salt structure. In our phonon calculations [not shown] for the wurtzite and zinc-blende structures, both are dynamically stable at 0 K. Our theoretical results thereby support the recent experimental findings of synthesized wurtzite and zinc-blende LaN~\cite{LaN_synthesis}.

	\begin{figure}
		\centering
		\includegraphics[width=\linewidth]{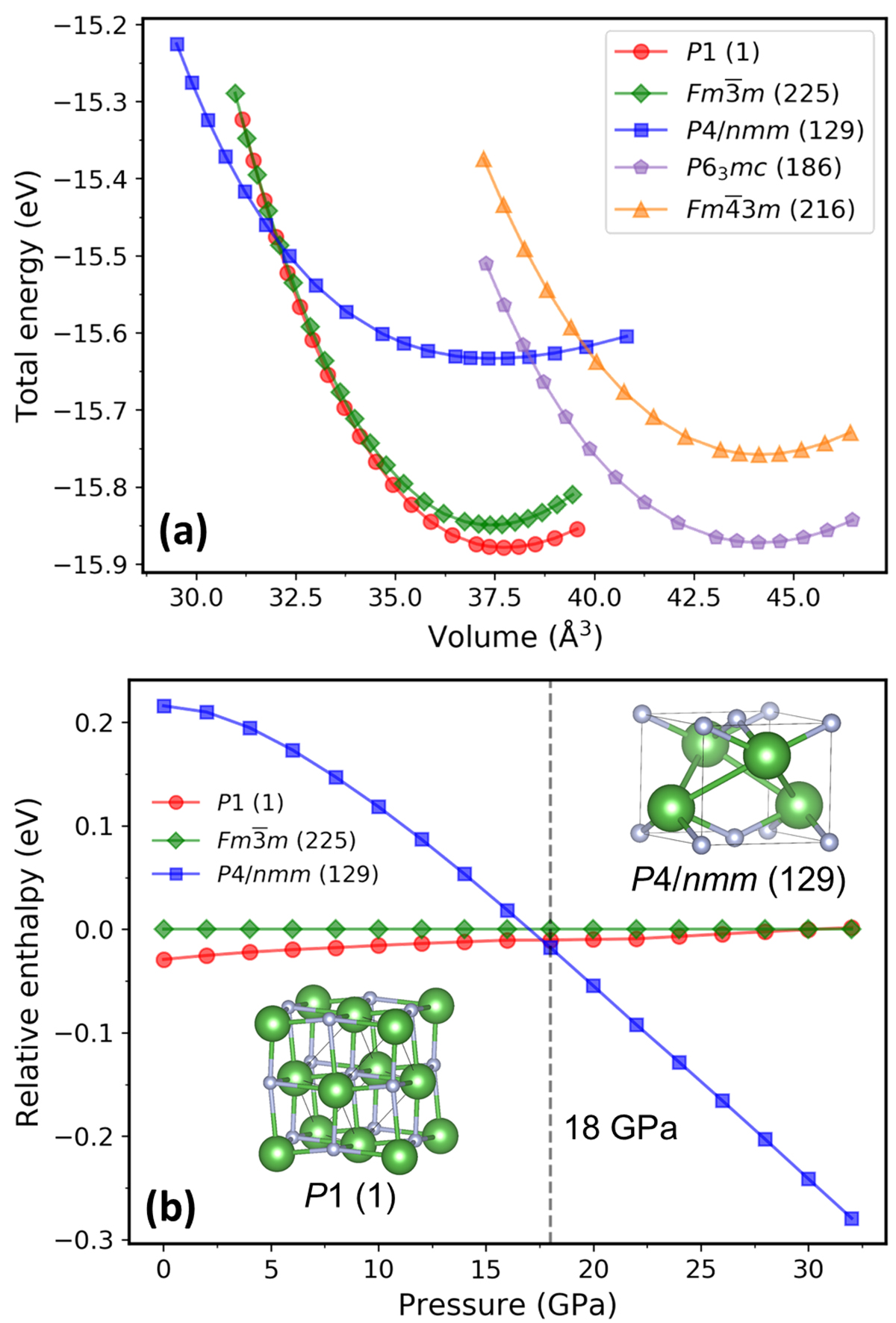}
		\caption {(a) Total-energy versus volume curves of LaN in different crystal symmetries, including the low-symmetry structure $P1$ (space group No. 1), rock salt $Fm\bar{3}m$ (No. 225), PbO-like tetragonal structure $P4/nmm$ (No. 129), wurtzite $P6_3mc$ (No. 186), and zinc blende $Fm\bar{4}3m$ (No. 216). (b) Relative formation enthalpy as a function of pressure using rock-salt LaN as the reference. The result indicates a pressure-induced structural phase transition from $P1$ to $P4/nmm$ symmetry near 18 GPa.}
		\label{Fig.3}
	\end{figure}
	
	Figure 3(a) also shows that the total energy-volume curves of $P1$ and $P4/nmm$ intersect around 32 \r{A}$^3$. Therefore, applying external pressure can favor the PbO-like $P4/nmm$ tetragonal structure. Figure 3(b) shows the calculated relative formation enthalpy as a function of pressure, using the rock-salt $Fm\bar{3}m$ structure as the reference. The result indicates that when the external pressure is below 18 GPa, $P1$-LaN has the lowest enthalpy. Above 18 GPa, $P4/nmm$-LaN becomes the lowest enthalpy structure. In other words, our DFT calculations suggest a pressure-induced structural phase transition from $P1$ to $P4/nmm$ symmetry in LaN near 18 GPa at 0~K. Our predicted transition pressure is similar to that from a previous theoretical study by Mukherjee {\it et al.}, who proposed a transition pressure of 19 GPa for LaN to transit from $Fm\bar{3}m$ to $P4/nmm$ symmetry~\cite{LaN_pressure_theory}. The theory results are also consistent with the experimental finding by Schneider {\it et al.}, who observed the $Fm\bar{3}m$ structure in ambient conditions, and it undergoes a phase transition to the PbO-like $P4/nmm$ tetragonal structure at $\sim 22.8$ GPa~\cite{LaN_pressure_exp}.
	
	\begin{figure}
		\centering
		\includegraphics[width=\linewidth]{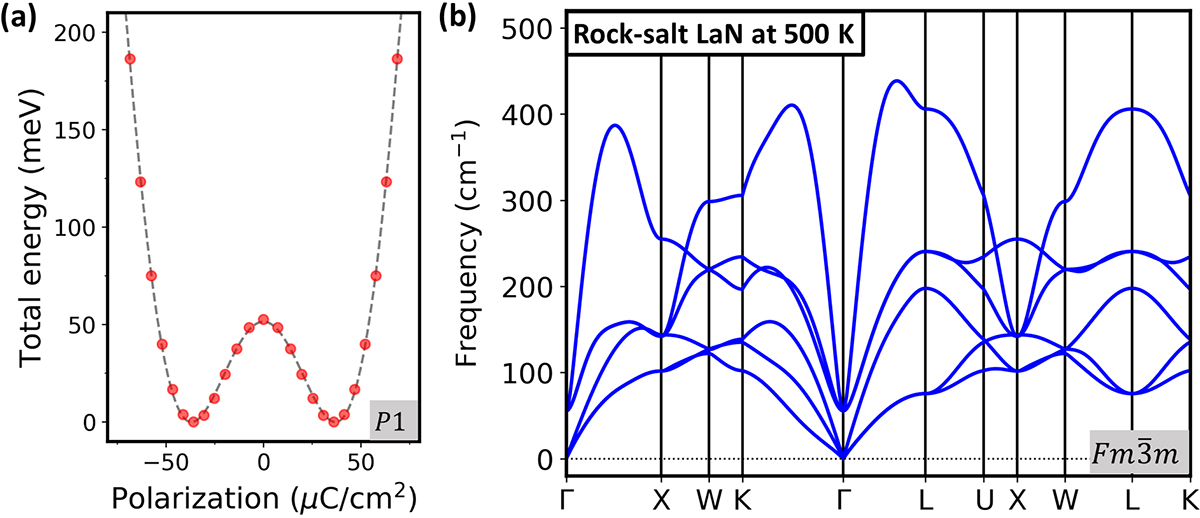}
		\caption{
		(a) Double-well potential energy versus polarization in $P1$-LaN, using the ground state energy as a reference. The ground state shows a spontaneous polarization of 36 $\mu$C/cm$^2$. The gray dashed line is obtained by fitting the energy with a 6th-order expansion in the order parameter.
		(b) Renormalized phonon dispersion of rock-salt ($Fm\bar{3}m$) LaN at 500 K. Compared to the phonon dispersion at 0 K in Fig. 1(b), the absence of imaginary mode here shows that rock-salt LaN can be stabilized at high temperature.}
		\label{Fig.4}
	\end{figure}
	
		\begin{figure*}[ht!]
		\includegraphics[width=\textwidth]{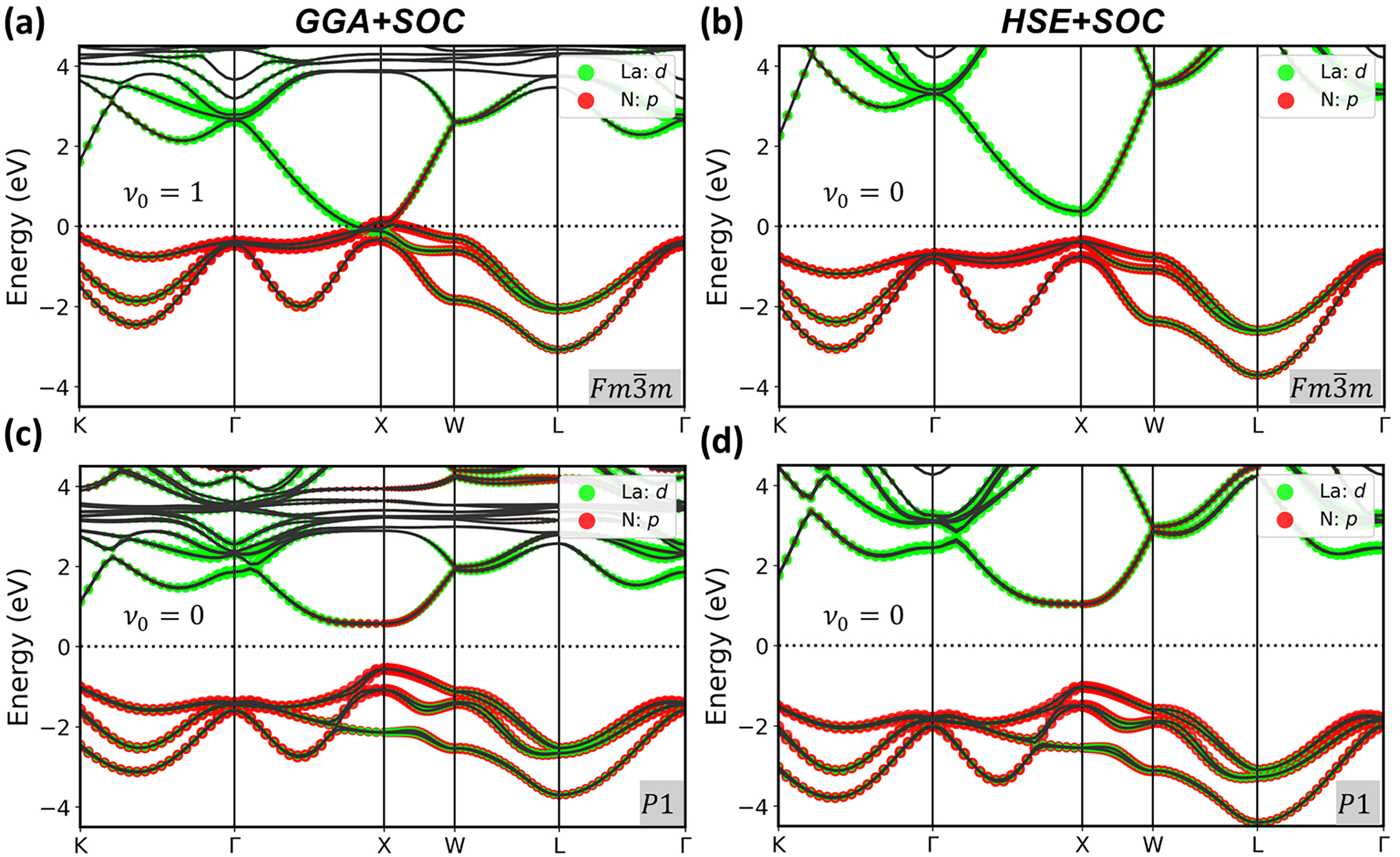}
		\caption{Electronic band structures with spin-orbit coupling (SOC) and $\mathbb{Z}_2$ topological indices for rock-salt LaN, calculated with (a) the PBE-GGA functional and (b) the HSE06 hybrid functional. (c)-(d) are similar plots as (a)-(b), but for $P1$ LaN. The labelings of high-symmetry points are based on the fcc lattice. The Fermi level (horizontal dotted line) is set to 0 eV. GGA typically underestimates the band gap, which can be corrected by a more advanced hybrid functional.}
		\label{Fig.5}
	\end{figure*}
	
	On the other hand, although our DFT calculation predicts a stable $P1$-LaN at 0 K, this structure has not been reported before. As mentioned above, the rock-salt $Fm\bar{3}m$-LaN is generally acknowledged in the literature as the correct structure in ambient conditions. These results imply that LaN may exhibit a temperature-induced structural transition from $P1$ to $Fm\bar{3}m$ symmetry. To confirm our conjecture that rock-salt LaN can be stabilized at high temperature, we calculate its finite-temperature (or renormalized) phonon dispersion. To include anharmonic effects, we use the TDEP method~\cite{TDEP} to obtain the effective 2nd-order force constants and fit them via AIMD simulations. Figure 4 shows the renormalized phonon dispersion of rock-salt LaN at 500 K. No imaginary mode exists, indicating the dynamical stability of rock-salt LaN at high temperature.
	
	In fact, the situation here for LaN is very similar to that in the ferroelectric materials GeSe and GeTe~\cite{GeSe,GeTe_Xia}. There, the structural transitions induced by temperature are usually accompanied by ferroelectric transitions. For example, rhombohedral $\alpha$-GeTe (space group No. 160, $R3m$ symmetry) undergoes a ferroelectric transition and a structural change to rock-salt $\beta$-GeTe (No. 225, $Fm\bar{3}m$) near 700 K~\cite{GeTe_Xia}.
An important signature of ferroelectric materials is the presence of a double-well potential energy surface, which can be obtained by the nudged elastic band method. Motivated by the similarity between GeTe and LaN, we compute the spontaneous electric polarization of $P1$-LaN based on the modern theory of polarization~\cite{modern_polarization}. Figure 4(a) shows the potential energy as a function of polarization, which exhibits a double-well energy landscape. In our calculations, $P1$-LaN has a switching barrier of 53 meV per formula unit, and the ground state structure displays a net polarization of 36 $\mu$C/cm$^2$. Therefore, LaN is likely to undergo a concurrent ferroelectric transition and structural change from $P1$ to $Fm\bar{3}m$ symmetry when the temperature is increased.
	
	Here, we do not attempt to discuss the exact critical transition temperature, because it is computationally demanding to calculate and also depends on the chosen DFT functional. In particular, our phonon dispersions computed with the PBE functional still contain negative frequencies of -60 and -44 $cm^{-1}$ at the $\Gamma$ point at temperatures 100 K and 300 K, respectively. We have also calculated the phonon dispersion at 300 K using the LDA functional, and the result shows that the negative phonon frequency at 300 K is reduced to only -19 $cm^{-1}$.

	We next shift our focus to electronic and topological properties. In a 3D system with time reversal symmetry (TRS), band topology can be characterized by the $\mathbb{Z}_2$ indices ($\nu_0$; $\nu_1, \nu_2, \nu_3$), if an energy gap exists at each $k$-point in the BZ~\cite{fu2007}. A strong topological insulator (STI) has a strong index $\nu_0$ = 1 with an odd number of Dirac cones on any surface. In contrast, the weak TI phase has $\nu_0$ = 0, and it can be regarded as a stacking of 2D quantum spin Hall~\cite{maciejko2011} layers along the ($\nu_1,\nu_2,\nu_3$) ``mod 2'' reciprocal lattice vector~\cite{fu2007}. Therefore, the weak indices ($\nu_1,\nu_2,\nu_3$) depend on the choice of unit cell, and here they are defined by the primitive crystal structure. To understand the $\mathbb{Z}_2$ topology of a system with both TRS and inversion symmetries~\cite{fu2007b}, one can study simply the product of the parity eigenvalues of the occupied bands at the time-reversal invariant momenta (TRIM). In the rock-salt structure, the TRIM include 1 $\Gamma$, 3 $X$, and 4 $L$, as shown in Fig. 2(b). Therefore, only a sign change in the parity eigenvalue at $\Gamma$ or $X$ point can lead to a change in $\nu_0$ (i.e. a STI topological phase transition). On the other hand, the three weak indices depend on 2 (even number) $X$ and 2 (even number) $L$ points, and thereby they are always 0. 
	

	\begin{figure*}
		\includegraphics[width=\textwidth]{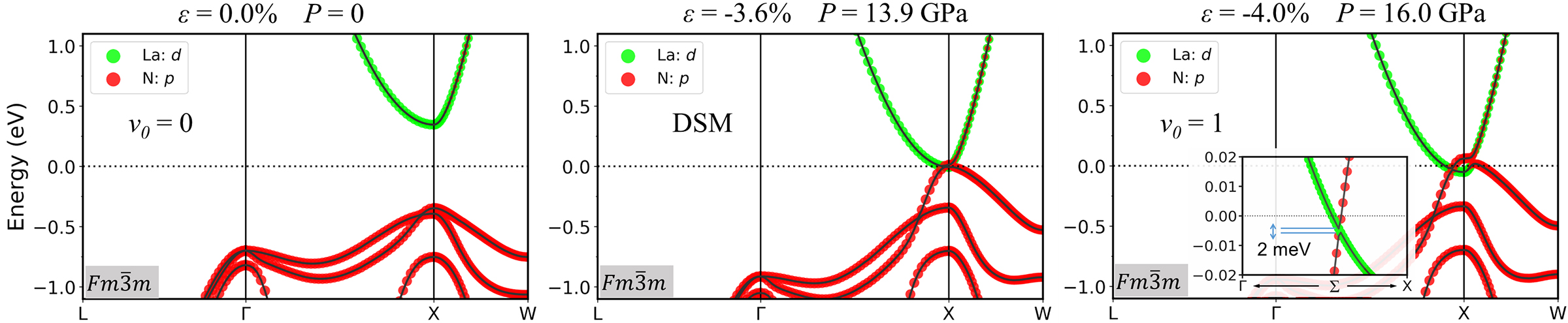}
		\caption{Electronic band structures and $\mathbb{Z}_2$ topological indices of rock-salt LaN under different strain $\varepsilon$ and corresponding pressure $P$. Left: $P$ = 0, trivial insulator ($\nu_0$ = 0). Middle: $P$ = 13.9 GPa, Dirac semimetal (DSM). Right: $P$ = 16.0 GPa, strong topological insulator ($\nu_0$ = 1). The results show a pressure-driven topological phase transition in rock-salt LaN.}
		\label{Fig.6}
	\end{figure*}

	The electronic band topology of LaX was first investigated by Zeng \textit{et al.} using the DFT+U method~\cite{LaX_TI_Lin}, which however does not fix the issue of underestimating LaX's band gaps and may lead to incorrect topological classifications. Later DFT studies based on the modified Becke-Johnson (mBJ) meta-GGA functional~\cite{mBJ_1, mBJ_2} or screened hybrid functionals~\cite{HSE06} have addressed this issue, and thereby their topological classifications are potentially more appropriate~\cite{LaBi_LaSb_mBJ_guo2016,LaAs_mBJ_XMR_yang2017, LaBi_LaSb_topo_dey2018}. Here, we study the electronic band structures using both PBE-GGA~\cite{PBE} and HSE06~\cite{HSE06}. For rock-salt LaN, the GGA band structure in Fig. 5(a) shows a semimetal and a topologically non-trivial STI phase ($\nu_0 = 1$) with band inversion at the $X$ point. However, this STI assignment is potentially incorrect, due to an underestimation of the overall gap between valence and conduction bands. In contrast, the more advanced HSE06 band structure in Fig. 5(b) shows that rock-salt LaN is a semiconductor with a direct band gap $\sim 0.8$ eV at the $X$ point, and the material is topologically trivial ($\nu_0 = 0$). We also calculate the band structures of $P1$-LaN. As shown in Figs. 5(c)-(d), the PBE-GGA and HSE06 results both show a semiconducting phase with a direct band gap $\sim 1.1$ and 2.2 eV, respectively. The orbitally-resolved band structures of $P1$-LaN show contributions from nitrogen's $p$-orbitals near the conduction band minimum. However, we note that this apparent ``band inversion" at the $X$ point is not related to SOC. Instead, the orbital hybridization is caused by the lower-symmetry of the distorted $P1$ structure. Due to the absence of inversion symmetry, the topological nature cannot be simply determined from the parity eigenvalues of the TRIM. Instead, here $\nu_0$ is calculated by tracking directly the Wannier charge centers, and $P1$-LaN is found to be topologically trivial ($\nu_0 = 0$).

	Finally, we study pressure-induced topological phase transitions (TPTs) in LaX. Recently, researchers have reported that the band topology of LaAs and LaSb can be controlled by pressure~\cite{LaAs_hydrostatic_pressure,  LaSb_topo_transition_HSE_guo2017}. External pressure can modify the band gap, leading to a band inversion between valence and conduction bands. Following a similar idea, here we calculate the HSE06 band structures of LaN under external pressure. As shown in the left panel of Fig. 6, the unstrained (ambient-condition) rock-salt LaN is topologically trivial ($\nu_0$ = 0). When a hydrostatic pressure of 13.9 GPa is applied, the energy gap is closed at the $X$ point, and rock-salt LaN becomes a Dirac semimetal (middle panel of Fig. 6). When the external pressure is above 13.9 GPa, a band inversion occurs (right panel of Fig. 6), and the material becomes a STI ($\nu_0$ = 1). Since inversion is preserved across the transition, the Dirac semimetal appears exactly and only at the TPT, and must necessarily appear at a TRIM~\cite{murakami2007,murakami2008}, here the three $X$ points in the BZ. We note that in the non-trivial STI phase, there is a very small gap $\sim$ 2 meV (the $\Sigma$ point in the insert, right panel of Fig. 6) due to SOC. We also explore possible TPTs in $P1$-LaN. However, for external pressure less than 18 GPa, the band gap remains finite, which means that $P1$-LaN is always a trivial insulator before transforming into the PbO-like $P4/nmm$ tetragonal structure.	
	

	\section{\label{sec:level1}Conclusion}
	Using density functional theory and evolutionary crystal searches, we have discovered a new, low-temperature structure of LaN belonging to space group No. 1 with the $P1$ symmetry. This lower-symmetry phase, which can be regarded as a distorted rock-salt structure, displays ferroelectricity with a spontaneous polarization of 36 $\mu$C/cm$^2$. Using {\it ab initio} molecular dynamics, we found that when the temperature is increased, the $P1$-LaN will undergo simultaneous ferroelectric and structural transitions to the more symmetric rock-salt structure (space group 225, $Fm\bar{3}m$). In our enthalpy calculations, $P1$-LaN also could undergo a pressure-induced phase transition to the PbO-like $P4/nmm$ tetragonal structure, when the external pressure is above 18 GPa. In our study of topological properties, the high-temperature rock-salt phase of LaN was found to undergo a pressure-induced transition from a trivial insulator to a strong topological insulator near 13.9 GPa. By increasing the temperature, we also have predicted another topological and structural transition from $P1$-LaN to rock-salt LaN, when the pressure is between 13.9 and $\sim 18$ GPa. These results demonstrate that LaN exhibits a rich phase diagram containing different ferroelectric, structural, and topological phases, and lanthanum monopnictides are intriguing quantum materials that can serve as a platform for exploring novel phase transitions induced by temperature and pressure.

\section*{Declaration of Competing Interest}
The authors declare that they have no known competing financial interests or personal relationships that could have appeared to influence the work reported in this paper.

\section*{Data Availability Statement}
The data that support the findings of this study are available on request from the corresponding author.

\section*{Acknowledgments}
The calculations were performed on the Frontera computing system at the Texas Advanced Computing Center. Frontera is made possible by NSF award OAC-1818253. J.M. was supported by NSERC Discovery Grants Nos. RGPIN-2020-06999 and RGPAS-2020-00064; the CRC Program; CIFAR; a Government of Alberta MIF Grant; a Tri-Agency NFRF Grant (Exploration Stream); and the PIMS CRG program.
	
		\begin{figure*}
		\includegraphics[width=\textwidth]{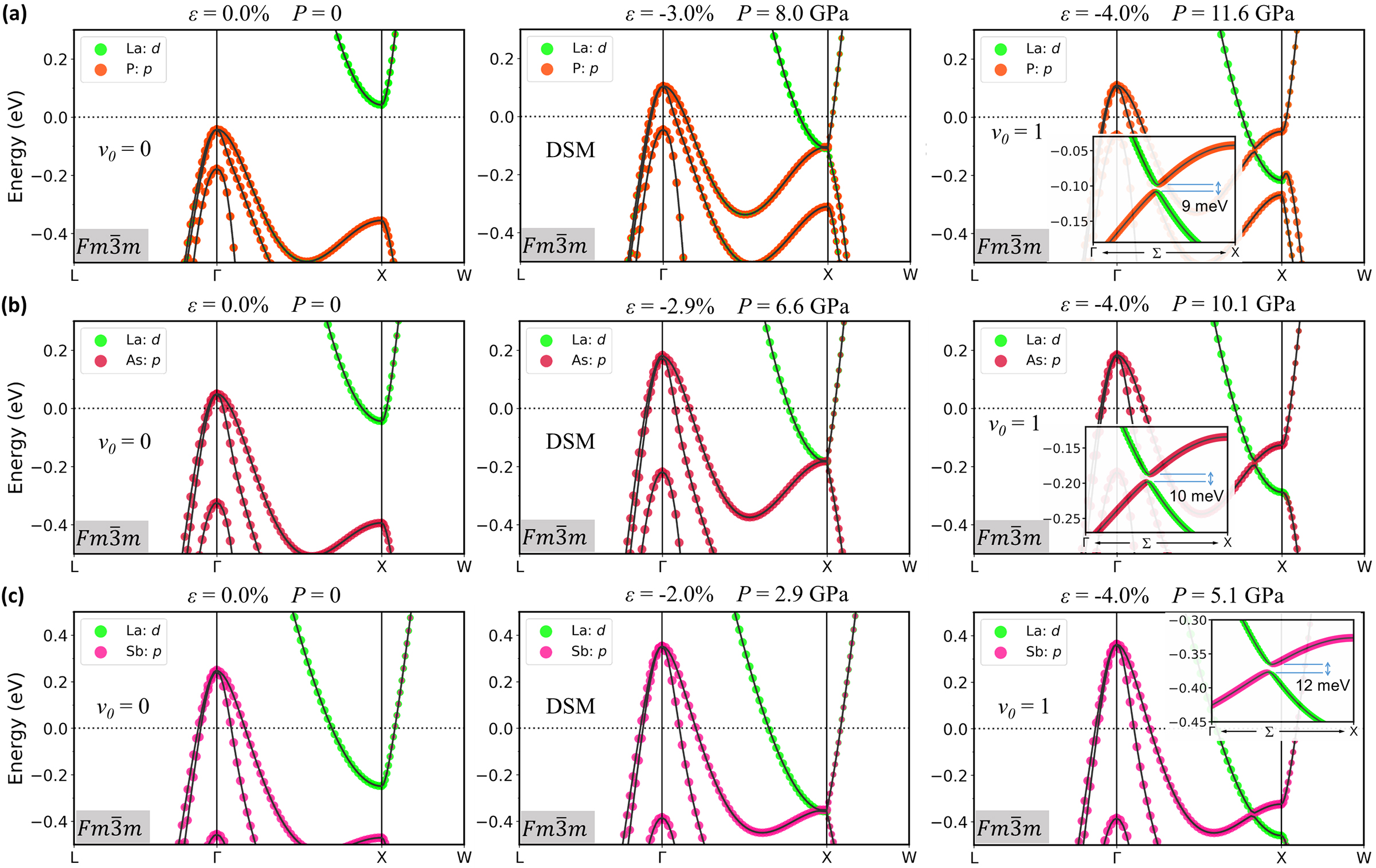}
		\caption{
		Electronic band structures and $\mathbb{Z}_2$ topological indices of rock-salt (a) LaP, (b) LaAs, and (c) LaSb, under different strain $\varepsilon$ and corresponding pressure $P$. Left: Trivial insulator ($\nu_0$ = 0). Middle: Dirac semimetal (DSM). Right: Strong topological insulator ($\nu_0$ = 1). The results show pressure-driven topological transitions in rock-salt lanthanum monipnictides.
		}
		\label{Fig.7}
	\end{figure*}

	\appendix
	\counterwithin{figure}{section}
	\section{}	
		
	In this Appendix, we study pressure-induced topological phase transitions in rock-salt LaP, LaAs, and LaSb, by using a hybrid density functional. The electronic band structures of unstrained LaP, LaAs, and LaSb are shown in the left panels of Fig. 7, and they are all found to be topologically trivial ($\nu_0 = 0$) at zero pressure $P=0$. The middle panels of Fig. 7 show the band structures of LaP, LaAs, and LaSb at the topological transition (Dirac semimetal), where the critical hydrostatic pressures are $P = 8.0$ GPa (with strain value $\varepsilon$ = -3.0\%), 6.6 GPa ($\varepsilon$ = -2.9\%), and 2.9 GPa ($\varepsilon$ = -2.0\%), respectively. Finally, the right panels of Fig. 7 display the high-pressure band structures, showing band gaps at the $\Sigma$ point with gap sizes of 9, 10, and 12 meV, respectively for LaP, LaAs, and LaSb. This increasing gap size is expected with an increasing spin-orbital coupling strength as one goes down along the periodic table.

	\bibliography{bibfile}
		
	\end{document}